\documentclass[aps,preprintnumbers,prl,twocolumn,superscriptaddress]{revtex4}

\usepackage{epsfig,latexsym,cancel,amssymb,amsmath}
\usepackage{graphicx}
\usepackage{feynmp}
\usepackage{subfigure}
\usepackage{epstopdf}
\usepackage{color}

\def\be{\begin{equation}}
\def\ee{\end{equation}}
\def\bea{\begin{eqnarray}}
\def\eea{\end{eqnarray}}

\def\bbuildrel#1_#2^#3{\mathrel{\mathop{\kern 0pt#1}\limits_{#2}^{#3}}}
\def\slash#1{\setbox0=\hbox{$#1$}#1\hskip-\wd0\dimen0=5pt\advance
       \dimen0 by-\ht0\advance\dimen0 by\dp0\lower0.5\dimen0\hbox
         to\wd0{\hss\sl/\/\hss}}

\newcommand{\gae}{\lower 2pt \hbox{$\, \buildrel {\scriptstyle >}\over {\scriptstyle
\sim}\,$}}
\newcommand{\lae}{\lower 2pt \hbox{$\, \buildrel {\scriptstyle <}\over {\scriptstyle
\sim}\,$}}

\newcommand{\beq}{\begin{eqnarray}}
\newcommand{\eeq}{\end{eqnarray}}

\newcommand{\ba}{\begin{array}}
\newcommand{\ea}{\end{array}}

\long\def\symbolfootnote[#1]#2{\begingroup%
\def\thefootnote{\fnsymbol{footnote}}\footnote[#1]{#2}\endgroup}

\newcommand{\chpt}{$\chi$PT\ }

\def\lsim{\mathrel{\rlap{\lower4pt\hbox{\hskip1pt$\sim$}}
    \raise1pt\hbox{$<$}}}         
\def\gsim{\mathrel{\rlap{\lower4pt\hbox{\hskip1pt$\sim$}}
    \raise1pt\hbox{$>$}}}         

\def\lsim{\:\raisebox{-0.5ex}{$\stackrel{\textstyle<}{\sim}$}\:}
\def\gsim{\:\raisebox{-0.5ex}{$\stackrel{\textstyle>}{\sim}$}\:}

\def\issue(#1,#2,#3){{\bf #1}, #2 (#3)}
\def\opcit(#1){ {\em op. cit.}, #1}

\def\APP(#1,#2,#3){Acta Phys.\ Polon.\ \issue(#1,#2,#3)}
\def\ARNPS(#1,#2,#3){Ann.\ Rev.\ Nucl.\ Part.\ Sci.\ \issue(#1,#2,#3)}
\def\CPC(#1,#2,#3){Comp.\ Phys.\ Comm.\ \issue(#1,#2,#3)}
\def\CIP(#1,#2,#3){Comput.\ Phys.\ \issue(#1,#2,#3)}
\def\EPJC(#1,#2,#3){Eur.\ Phys.\ J.\ C\ \issue(#1,#2,#3)}
\def\EPJD(#1,#2,#3){Eur.\ Phys.\ J. Direct\ C\ \issue(#1,#2,#3)}
\def\IEEETNS(#1,#2,#3){IEEE Trans.\ Nucl.\ Sci.\ \issue(#1,#2,#3)}
\def\IJMP(#1,#2,#3){Int.\ J.\ Mod.\ Phys. \issue(#1,#2,#3)}
\def\JHEP(#1,#2,#3){J.\ High Energy Physics \issue(#1,#2,#3)}
\def\JPG(#1,#2,#3){J.\ Phys.\ G \issue(#1,#2,#3)}
\def\MPL(#1,#2,#3){Mod.\ Phys.\ Lett.\ \issue(#1,#2,#3)}
\def\NP(#1,#2,#3){Nucl.\ Phys.\ \issue(#1,#2,#3)}
\def\NIM(#1,#2,#3){Nucl.\ Instrum.\ Meth.\ \issue(#1,#2,#3)}
\def\PL(#1,#2,#3){Phys.\ Lett.\ \issue(#1,#2,#3)}
\def\PRD(#1,#2,#3){Phys.\ Rev.\ D \issue(#1,#2,#3)}
\def\PRL(#1,#2,#3){Phys.\ Rev.\ Lett.\ \issue(#1,#2,#3)}
\def\SJNP(#1,#2,#3){Sov.\ J. Nucl.\ Phys.\ \issue(#1,#2,#3)}
\def\ZPC(#1,#2,#3){Zeit.\ Phys.\ C \issue(#1,#2,#3)}

\def\beq{\begin{equation}}
\def\eeq{\end{equation}}
\def\bea{\begin{eqnarray}}
\def\eea{\end{eqnarray}}
\def\to{\rightarrow}

\begin{document}


\preprint{
\vbox{
\hbox{ACFI-T13-01,\, CALT CALT 68-2859} 
}}

\title{The Muon Anomalous Magnetic Moment and the Pion Polarizability}
\author{Kevin T. Engel}
\affiliation{California Institute of Technology, Pasadena, CA 91125 USA}
\author{Michael J. Ramsey-Musolf}
\affiliation{Physics Department, University of Massachusetts Amherst, Amherst, MA 01003, USA\\ and Kellogg Radiation Laboratory, California Institute of Technology, Pasadena, CA 91125}

\begin{abstract} 
We compute the charged pion loop contribution to the muon anomalous magnetic moment $a_\mu$, taking into account the effect of the charged pion polarizability, $(\alpha_1-\beta_1)_{\pi^+}$. We evaluate this contribution using two different models that are consistent with the requirements of chiral symmetry in the low-momentum regime and perturbative quantum chromodynamics in the asymptotic region. The result increases the disagreement between the present experimental value for $a_\mu$ and the theoretical, Standard Model prediction by as much as $\sim 60\times 10^{-11}$, depending on the value of $(\alpha_1-\beta_1)_{\pi^+}$ and the choice of the model. The planned determination of $(\alpha_1-\beta_1)_{\pi^+}$ at Jefferson Laboratory will eliminate the dominant parametric error, leaving a theoretical model uncertainty commensurate with the error expected from planned Fermilab measurement of $a_\mu$.


\end{abstract}

\maketitle

The measurement of the muon anomalous magnetic moment, $a_\mu$, provides one of the most powerful tests of the Standard Model of particle physics and probes of physics that may lie beyond it. The present experimental value obtained by the  E821 Collaboration\cite{Bennett:2006fi,Bennett:2004pv,Bennett:2002jb}  $a_\mu^\mathrm{exp}= 116592089(63)\times 10^{-11}$ disagrees with the most widely quoted theoretical SM predictions  by $3.6\sigma$:  $a_\mu^\mathrm{SM}= 116591802(49)\times 10^{-11}$ (for recent reviews, see Ref.~\cite{Passera:2010ev,Jegerlehner:2009ry} as well as references therein). This difference may point to physics beyond the Standard Model (BSM) such as weak scale supersymmetry or very light, weakly coupled neutral gauge bosons\cite{Stockinger:2006zn,Hertzog:2007hz,Czarnecki:2001pv,Pospelov:2008zw} . A next generation experiment planned for Fermilab would reduce the experimental uncertainty by a factor of four\cite{FERMILAB-PROPOSAL-0989}. If a corresponding reduction in the theoretical, SM uncertainty were achieved, the muon anomalous moment could provide an even more powerful indirect probe of BSM physics.

The dominant sources of theoretical uncertainty are associated with non-perturbative strong interaction effects that enter the leading order hadronic vacuum polarization (HVP) and the hadronic light-by-light (HLBL) contributions: $\delta a_\mu^\mathrm{HVP}(\mathrm{LO}) = \pm 42 \times 10^{-11}$ and $\delta a_\mu^\mathrm{HLBL} = \pm 26 \times 10^{-11}$ \cite{Prades:2009tw}  (other authors give somewhat different error estimates for the latter \cite{Kinoshita:1984it,Bijnens:2001cq,Bijnens:1995cc,Bijnens:1995xf,Hayakawa:1997rq,Hayakawa:1996ki,Hayakawa:1995ps,Knecht:2001qg,Knecht:2001qf,Melnikov:2003xd,Bijnens:2007pz,Nyffeler:2009tw,RamseyMusolf:2002cy,Goecke:2011pe,Blum:2013qu} , but we will refer to these numbers as points of reference; see \cite{Nyffeler:2010rd} for a review). In recent years, considerable scrutiny has been applied to the determination of $a_\mu^\mathrm{HVP}(\mathrm{LO})$ from data on  $\sigma(e^+e^-\to\mathrm{hadrons})$ and hadronic $\tau$ decays.  A significant reduction in this HVP error will be needed if the levels of theoretical and future experimental precision are to be comparable. 

In this Letter, we concentrate on the more theoretically-challenging  $a_\mu^\mathrm{HLBL}$. At leading order in the expansion of the number of colors $N_C$, $a_\mu^\mathrm{HLBL}$ is generated by the pseudoscalar pole contributions that in practice turn out to be numerically largest. The contribution arising from charged pion loops is subleading in $N_C$, yet the associated error is now commensurate with the uncertainty typically quoted for the pseudoscalar pole terms. Both uncertainties are similar in magnitude to the goal experimental error for the proposed Fermilab measurement. Thus, it is of interest to revisit previous computations of the charged pion loop contribution, scrutinize the presently quoted error, and determine how it might be reduced. 

In previously reported work\cite{Engel:2012xb}, we completed a step in this direction by computing the amplitude $\Pi^{\mu\nu\alpha\beta}$  for light-by-light scattering for low-momentum off-shell photons. In this regime, Chiral Perturbation Theory  (\chpt) provides a first principles, effective field theory description of strong interaction dynamics that incorporates the approximate chiral symmetry of quantum chromodynamics (QCD) for light quarks. Long-distance hadronic effects can be computed order-by-order in an expansion of $p/\Lambda_\chi$, where $p$ is a typical energy scale (such as the pion mass $m_\pi$ or momentum) and $\Lambda_\chi=4\pi F_\pi\sim 1$ GeV is the hadronic scale with $F_\pi=93.4$ MeV being the pion decay constant. At each order in the expansion, presently incalculable strong interaction effects associated with energy scales of order $\Lambda_\chi$ are parameterized by a set of effective operators with {\em a priori} unknown coefficients. After renormalization, the finite parts of these coefficients -- \lq\lq low energy constants" (LECs) -- are fit to experimental results and then used to predict other low-energy observables. 

Working to next-to-next-to leading order (NNLO) in this expansion, we showed that models used to date in computing the full charged pion contribution to $a_\mu^\mathrm{HLBL}$ do not reproduce the structure of the low-momentum off-shell HLBL scattering amplitude implied by the approximate chiral symmetry of QCD. In particular, these models fail to generate terms in the amplitude proportional to the pion polarizability, a $\pi\pi\gamma\gamma$ interaction arising from two terms in the $\mathcal{O}(p^4)$ chiral Lagrangian:
\bea
\label{eq:leff1}
\mathcal{L} & \supset & ie \alpha_9\ F_{\mu\nu}\, \mathrm{Tr}\ \left(Q\left[D^\mu\Sigma, D^\nu\Sigma^\dag\right]\right)\\
\nonumber
&&+e^2\alpha_{10}\ F^2\ \mathrm{Tr}\left(Q\Sigma Q\Sigma^\dag\right)\ \ \ ,
\eea
where $Q=\mathrm{diag}(2/3,-1/3)$ is the electric charge matrix and $\Sigma=\mathrm{exp}(i \tau^a\ \pi^a/F_\pi)$ with $a=1,2,3$ giving the non-linear realization of the spontaneously broken chiral symmetry. The finite parts of the coefficients, $\alpha_9^r$ and $\alpha_{10}^r$, depend on the renormalization scale, $\mu$. 

The first term in Eq.~(\ref{eq:leff1}) gives the dominant contribution to the pion charge radius for $\mu=m_\rho$, the $\rho$-meson mass. The polarizability amplitude arises from both terms and is proportional to the $\mu$-independent combination $\alpha_9^r+\alpha_{10}^r$.  An experimental value has been obtained from the measurement of the rate for radiative pion decay \cite{arXiv:0801.2482}, yielding $(\alpha_9^r+\alpha_{10}^r)_\mathrm{rad} = (1.32\pm 0.14)\times 10^{-3}$ (see also Refs.~\cite{Bijnens:2006zp,arXiv:0810.0760}). On the other hand, direct determinations of the polarizability $(\alpha_1-\beta_1)_{\pi^+}$ have been obtained  from radiative pion photoproduction $\gamma p\to\gamma^\prime\pi^+ n$ and the hadronic Primakov process $\pi A\to\pi^\prime \gamma A$ where $A$ is a heavy nucleus. Using
\be
\label{eq:amb}
(\alpha_1-\beta_1)_{\pi^+} = 8 \alpha (\alpha_9^r+\alpha_{10}^r)/(F_\pi^2 m_\pi)+\cdots\ \ \ ,
\ee
where the \lq\lq $+\cdots$" indicate corrections that vanish in the chiral limit (see {\em e.g.}, Refs.~\cite{Burgi:1996qi,Gasser:2006qa}), these direct measurements yield  $(\alpha_9^r+\alpha_{10}^r)_{\gamma p} = (3.1\pm 0.9)\times 10^{-3}$\cite{Ahrens:2004mg}  and   $(\alpha_9^r+\alpha_{10}^r)_{\pi A} = (3.6\pm 1.0)\times 10^{-3}$\cite{Antipov:1982kz}, respectively. The COMPASS experiment has undertaken a new determination  using the hadronic Primakov process\cite{compass}, while a measurement of the process $\gamma\gamma\pi^+\pi^-$  is underway at Frascati. A determination using the reaction $\gamma A\to \pi^+\pi^- A$ has been has been approved for the GlueX detector in Hall D at Jefferson Laboratory (JLab), with an anticipated absolute uncertainty of $0.16\times 10^{-4}$\cite{jlab}.

\begin{figure}
\subfigure[]{
\label{fig:lo}
\includegraphics[width=25mm,height=25mm]{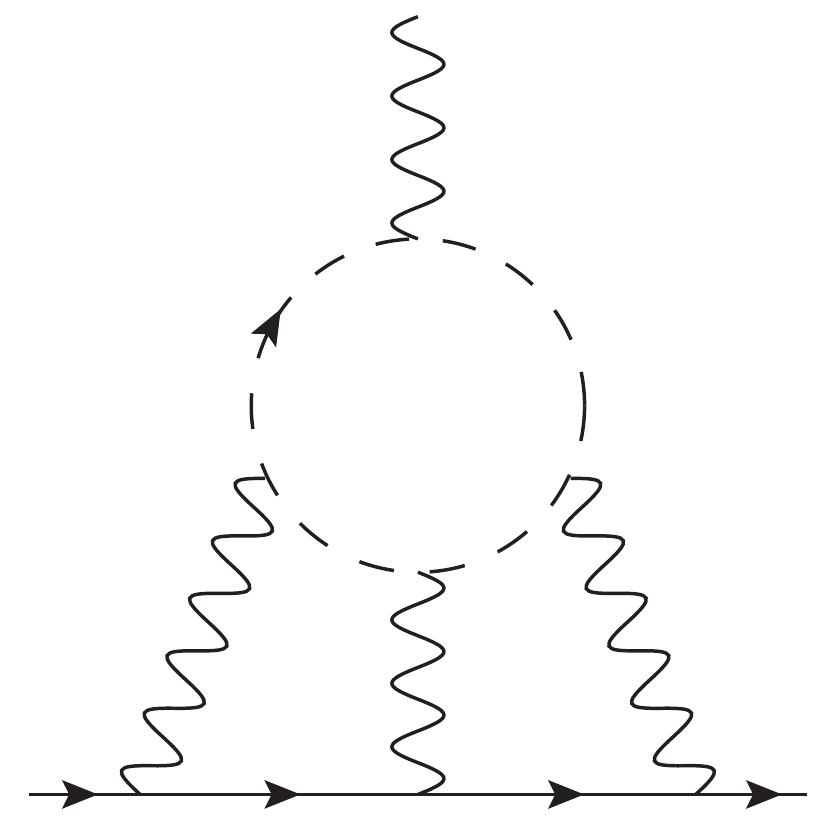}
}
\subfigure[]{
\label{fig:rho}
\includegraphics[width=25mm,height=25mm]{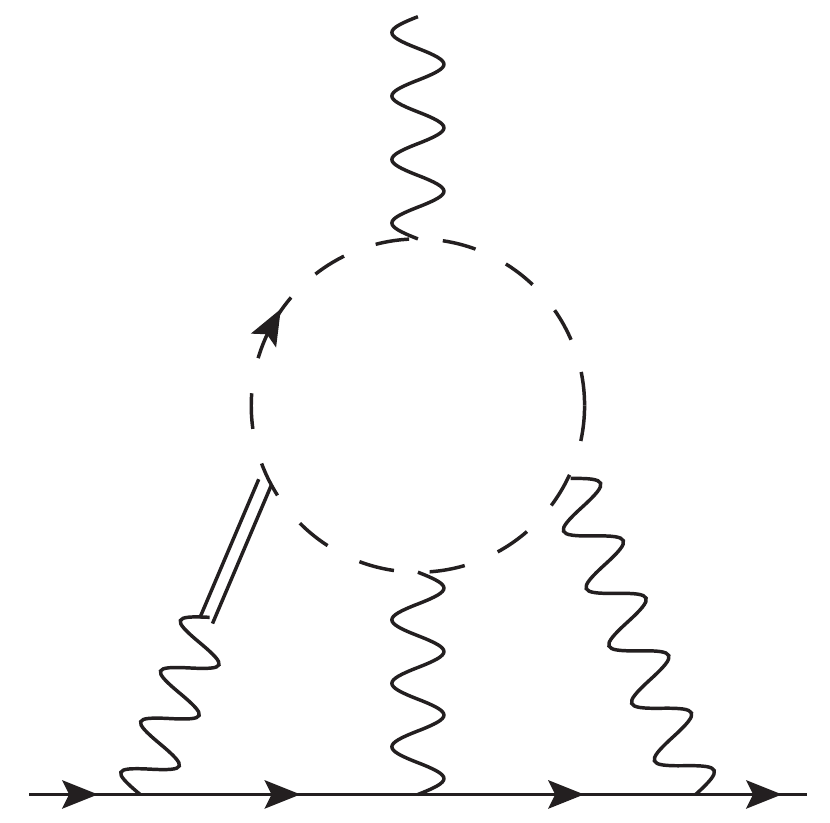}
}
\subfigure[]{
\label{fig:a1}
\includegraphics[width=25mm,height=25mm]{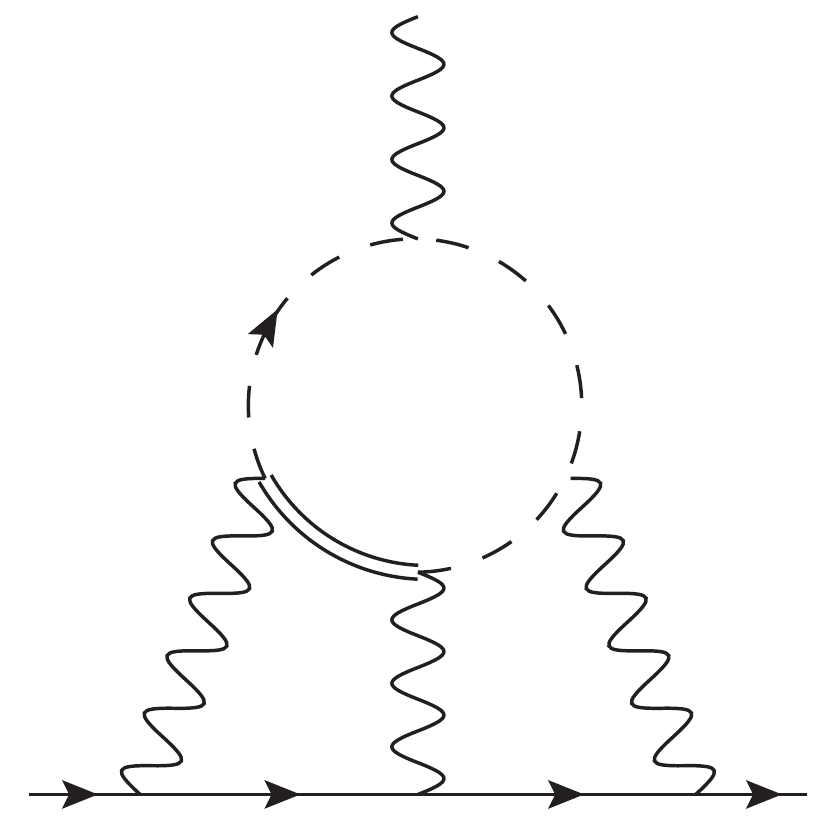}
}
\caption{Representative diagrams for charged pion loop contributions to $a_\mu^\mathrm{HLBL}$: (a) LO; (b) VMD ($\rho$-meson) model for the $\gamma\pi^+\pi^-$ vertex; (c) Model I and II $\gamma\gamma\pi^+\pi^-$ form factor.}
\end{figure}

In terms of the off-shell HLBL amplitude, the LO contribution arises solely from interactions appearing in the $\mathcal{O}(p^2)$ Lagrangian, corresponding to scalar quantum electrodynamics [see Fig.~\ref{fig:lo}]. The associated contribution to $a_\mu^\mathrm{HLBL}$ is finite and was first computed in Ref.~\cite{Kinoshita:1984it}. At NNLO in \chpt, one encounters distinct contributions to the low-momentum off-shell HLBL amplitude associated with the square of the pion charge radius
\be
r_\pi^2 = \frac{12}{F_\pi^2}\alpha_9^r(\mu)+\frac{1}{\Lambda_\chi^2}\left[ \ln\left(\frac{\mu^2}{m_\pi^2}\right)-1\right]\ \ \ 
\ee
and the polarizability LECs $\alpha_9^r+\alpha_{10}^r$. The resulting $\gamma\gamma\pi\pi$ vertex $V^{\mu\nu}\varepsilon_{\mu}(k_1)\varepsilon_{\nu}(k_2)$ for $|k_j^2| << m_\pi^2$ is given by
\bea
\nonumber
V^{\mu\nu}_{\chi\mathrm{PT}} & = & 2ie^2\Bigl\{g^{\mu\nu} + \frac{r_\pi^2}{6}\left[g^{\mu\nu} (k_1^2+k_2^2)-k_1^\mu k_1^\nu
-k_2^\mu k_2^\nu\right]\\
\label{eq:vmunu}
&&+\frac{4(\alpha_9+\alpha_{10})}{F_\pi^2}\left[k_1\cdot k_2 g^{\mu\nu}-k_2^\mu k_1^\nu\right]\Bigr\}\ \ \ .
\eea

From the model standpoint, efforts to incorporate the effects of pion substructure in electromagnetic interactions have generally followed a vector meson dominance (VMD) type of approach  [see Fig.~\ref{fig:rho}]. The first efforts with the simplest VMD implementation \cite{Kinoshita:1984it} were followed by  use an extended Nambu-Jona-Lasinio (ENJL) model\cite{Bijnens:1995xf}  the Hidden Local Symmetry (HLS) approach \cite{Hayakawa:1995ps,Hayakawa:1996ki}. In all cases, the associated contributions to the low-momentum off-shell HLBL amplitude match onto the \chpt\ results for the charge radius contributions when one identifies $r_\pi^2=6/m_\rho^2$. In contrast, the terms corresponding to $\alpha_9^r+\alpha_{10}^r$ are absent. Moreover, the coefficients of the polarizability contributions are comparable to those involving the charge radius, implying that the ENJL and HLS model results for the low-momentum regime are in  disagreement with the requirements of QCD. 

It is natural to ask whether this disagreement has significant implications for $a_\mu^\mathrm{HLBL}$. In an initial exploration of this question, the authors of Ref.~\cite{Bijnens:2012an,Abyaneh:2012ak} (see also \cite{Amaryan:2013eja})
 included the operators in Eq.~(\ref{eq:leff1}) in $\Pi^{\mu\nu\alpha\beta}(q, k_2, k_3,k_4)$, where $q$ and the $k_j$ are the real and virtual photon momenta, respectively, with $k_4=-(q+k_2+k_3)$.
Since the anomalous magnetic moment amplitude is linear in $q^\mu$, one need  retain only the first non-trivial term in an expansion in the external photon momentum. Differentiating the QED Ward identity $q_\lambda\Pi^{\lambda\nu\alpha\beta}=0$ with respect to $q^\mu$ implies that one may then express the HLBL amplitude entering the full integral for $a_\mu$ as\cite{Kinoshita:1984it}
\be
\Pi^{\mu\nu\alpha\beta}=-q_\lambda \frac{\partial\Pi^{\lambda\nu\alpha\beta}(q, k_2, k_3,-q-k_2-k_3)}{\partial q^\mu}\Bigr\vert_{q=0}\ \ \ .
\ee
Using this procedure, one finds that the contribution to $a_\mu$ proportional to $\alpha_9^r+\alpha_{10}^r$ is divergent. The authors of Refs.~\cite{Bijnens:2012an,Abyaneh:2012ak} thus regulated the integral by imposing a cutoff $K^2\equiv (k_2+k_3)^2 < (500\, \mathrm{MeV})^2$. The resulting impact on $a_\mu^\mathrm{HLBL}$ amounts to a $\sim 10\%$ increase in the magnitude of the overall charged pion loop contribution compared to the simplest VMD model prediction. 

Here, we report on a computation of $a_\mu^\mathrm{HLBL}$ that is consistent with the low-momentum requirements of QCD and that does not rely on an {\em ad hoc} cut off when extrapolating to the higher momentum regime where \chpt is not applicable. Instead, we employ two different models for the high-momentum behavior of the pion virtual Compton amplitude that are consistent with both the strictures of chiral symmetry in the low momentum region and the requirements of perturbative QCD in the domain of large photon virtuality. Compared with the conclusions of Refs.~\cite{Bijnens:2012an,Abyaneh:2012ak}, we find that the impact on $a_\mu$ may  be significant, leading to an increase in the discrepancy with the experimental result by as much as  $\sim 60\times 10^{-11}$, depending on which whether one takes the value of $\alpha_9^r+\alpha_{10}^r$ from radiative pion decays or direct determinations of the polarizability. The planned determination of $\alpha_9^r+\alpha_{10}^r$  at JLab could significantly reduce the spread of polarizability contributions to $a_\mu$. In the longer term, studies of the off-shell Compton amplitude could help reduce the theoretical uncertainty associated with interpolating between the chiral and asymptotic domains.

In modeling the higher momentum behavior of the polarizability, we are guided by several considerations:

\noindent{\em Chiral symmetry}. In the low-momentum regime, any model should reproduce the $\gamma\pi\pi$ and $\gamma\gamma\pi\pi$ interactions implied by the $\mathcal{O}(p^4)$ operators in Eq.~(\ref{eq:leff1}). As indicated earlier, neither the HLS nor the ENJL prescriptions are fully consistent with this requirement. While they incorporate the $\alpha_9^r$ (charge radius) contribution to the $\gamma\pi\pi$ interaction, they omit the contribution to the $\gamma\gamma\pi\pi$ interaction proportional to $\alpha_9^r+\alpha_{10}^r$.  

\noindent{\em Asymptotic behavior}. By using the operator product expansion, it is possible to show that the virtual Compton amplitude $T_{\mu\nu}(k, -k)$ must vanish as $1/k^2$ in the large $k^2$ regime. Neither the HLS nor the ENJL models satisfy this requirement. The HLS approach gives a non-vanishing $T_{\mu\nu}$ in the asymptotic limit, while in the ENJL framework the Compton amplitude falls off as $1/(k^2)^2$.

\noindent {\em Resonance saturation}. The LECs of the $\mathcal{O}(p^4)$ chiral Lagrangian are known to be saturated by spin-one meson resonances for $\mu\simeq m_\rho$. The ENJL and HLS approaches incorporate these \lq\lq resonance saturation" dynamics for the $\rho$-meson,  thereby obtaining the well-established relation $r_\pi^2=6/m_\rho^2$. By itself, however, inclusion of the $\rho$ does not lead to a correct description of the polarizability, as our study of the off-shell LBL amplitude demonstrates. 

\noindent{\em Pion mass splitting}. The degeneracy between charged and neutral pion masses is broken by the light quark mass difference and by the electromagnetic interaction. The pion polarizability and charge radius contribute to the latter when one embeds $T_{\mu\nu}$ in the one-loop pion self energy.  Retaining only the $\mathcal{O}(p^4)$ interactions in Eq.~(\ref{eq:leff1}) yields a divergent result that is rendered finite by inclusion of the operator\cite{Urech:1994hd}
\be
\label{eq:mpict}
\mathcal{L}  \supset \frac{e^2 C}{F_\pi^2}\, \mathrm{Tr}\, \left(Q\Sigma Q \Sigma^\dag\right)
\ee
that contributes $2 e^2 C/F_\pi^2$ to $m_{\pi^+}^2$ but nothing to $m_{\pi^0}^2$. Note that this contribution does not vanish in the chiral limit.
The  HLS approach also does not give a finite contribution. While it would be desirable that any model used to interpolate to the higher-momentum regime also reproduce the known value of $(\Delta m_\pi^2)_\mathrm{EM}$, this mass splitting does not enter directly into the HLBL amplitude to the order of interest here.  

One approach that satisfies the aforementioned criteria is to include the axial vector $a_1$ meson as well as the $\rho$ meson in the low-energy effective Lagrangian using the anti-symmetric tensor (AT) formulation\cite{Gasser:1984gg}. Detailed application of this approach to the pion polarizability and $(\Delta m_\pi^2)_\mathrm{EM}$ have been reported in Refs.~\cite{Donoghue:1993hj,Donoghue:1996zn}. The polarizability term in Eq.~(\ref{eq:vmunu}) is given by 
\be
\label{eq:axial}
4(\alpha_9^r+\alpha_{10}^r)_{a_1}= F_A^2/M_A^2\ \ \ ,
\ee
where $M_A$ and $F_A$ are the $a_1$ mass and electromagnetic coupling, respectively. (Consistency with a variety of theoretical and empirical considerations suggests taking $F_A=F_\pi$, which we follow below). Introduction of additional form factors leads to a finite contribution to $(\Delta m_\pi^2)_\mathrm{EM}$. Numerically, one finds that the experimental value for the pion mass splitting is well-reproduced.

Unfortunately, the $a_1$ AT model does not yield a finite result for $a_\mu^\mathrm{HLBL}$. We are, thus, motivated to consider alternative models that incorporate as many features of the $a_1$ dynamics as possible while satisfying the requirements of chiral symmetry, asymptotic scaling, and finite $a_\mu^\mathrm{HLBL}$.  Our strategy is to modify the $\gamma\gamma\pi\pi$ polarizability vertex by the introduction of vector meson-like form factors. We consider two models:
\be
\label{eq:model1}
\mathcal{L}_I = -\frac{e^2}{4} F_{\mu\nu} \pi^+ \left(\frac{1}{D^2+M_A^2}\right) F^{\mu\nu}\pi^- +\,  \mathrm{h.c.}\,  +\cdots\  ,
\ee
where $D_\mu=\partial_\mu+ieQ A_\mu$ is the covariant derivative and the $+\cdots$ are higher order terms in pion fields as dictated by chiral symmetry;  and
\be
\label{eq:model2}
\mathcal{L}_{II} = -\frac{e^2}{2 M_A^2} \pi^+\pi^- \left[\left(\frac{M_V^2}{\partial^2+M_V^2}\right) F^{\mu\nu}\right]^2+\cdots\ \ \ ,
\ee
with the partial derivatives acting only on the field strength tensors immediately to the right. In order to obtain the appropriate asymptotic behavior for $T_{\mu\nu}$, one must combine the Model I Lagrangian (\ref{eq:model1}) with either the AT or HLS formulation for the $\rho$-meson contributions, whereas in using the Model II Lagrangian (\ref{eq:model2}) it is necessary to employ the full VMD prescription for the $\rho$ (similar to the ENJL case, but with a momentum-independent $M_V$).  

By construction, both models reproduce the correct polarizability and  charge radius interactions that appear at $\mathcal{O}(p^4)$ and yield a Compton amplitude $T_{\mu\nu}(k, -k)$ that falls off as $1/k^2$. Both also generate a finite contribution to $a_\mu^\mathrm{HLBL}$. When  the Kawarabayashi-Suzuki-Fayyazuddin-Riazuddin (KSFR) relation\cite{Kawarabayashi:1966kd,Riazuddin:1966sw}
$M_A=\sqrt{2} M_V$ is imposed, 
Model I  gives rise to a finite contribution to $(\Delta m_\pi^2)_\mathrm{EM}$, whereas Model II requires the additional counterterm in Eq. (\ref{eq:mpict}). Since, however, $(\alpha_9^r+\alpha_{10}^r)\sim 1/M_A^2$ in these models [see Eq.~(\ref{eq:axial})], choosing  $M_A$ and $M_V$ to reproduce the experimental results for the polarizability and charge radius ($M_V=m_\rho$) may lead to a violation of the KSFR relation, thus implying for Model I both a divergent $(\Delta m_\pi^2)_\mathrm{EM}$ as well as incorrect asymptotic behavior for $T_{\mu\nu}$.  

An example of the additional diagrams needed for complete evaluation the new contributions to $a_\mu^\mathrm{HLBL}$ are shown in Fig.~\ref{fig:a1}. Note that in Model I, one encounters additional vertices associated with the action of the covariant derivative on the pion and field strength tensors. The form of the interactions in Models I and II facilitates numerical evaluation of the full $a_\mu^\mathrm{HLBL}$ integral.  The momentum space structures are propagator-like, thereby allowing us to employ conventional Feynman parameterization. In doing so, we follow the procedure described in Ref.~\cite{Kinoshita:1984it}, wherein evaluation of the loop integrals yields Feynman parameter integrals of the form
\be
\mathcal{M} = \int \Pi_j dx_j \delta(1-\sum x_j)\, \frac{N(x)}{U(x)^\alpha V(x)^\beta} \ \ \ .
\ee
Here, $U(x)$ and $V(x)$ arise from the denominator structure of the diagram, with $U(x)$ encoding the self-coupling of the loop momenta and $V(x)$ containing mass terms that govern the infrared behavior the integrand. The numerator $N(x)$ follows from the detailed structure of the interaction vertices. We have written a separate Monte Carlo routine for evaluating these Feynman parameter integrals, details of which will appear elsewhere. 

\begin{table}[hptb]
\caption{Charged pion loop contributions to $a_\mu^\mathrm{HLBL} $ in different approaches discussed in text. Second and third columns correspond to different values for the polarizability LECs, $(\alpha_9^r+\alpha_{10}^r)$: (a) $(1.32\pm 1.4)\times 10^{-3}$ and (b) $(3.1\pm 0.9)\times 10^{-3}$. Note that only the NLO/cut-off and Models I and II depend on these LECs. 
 } \label{tab:results}
\begin{tabular}{| c | c | c |}
\hline
Approach &  $a_\mu^{\pi^+\pi^-} \times 10^{11}$ (a) & $a_\mu^{\pi^+\pi^-} \times 10^{11}$ (b) \\ \hline
LO & -44 & -44 \\
HLS & -4.4 (2) & -4.4 (2) \\
ENJL & -19 (13) &  -19(13) \\
NLO/cut-off & -20 (5) & -24 (5) \\
Model I  & -11 & -34\\
Model II   & -40 & -71 \\
 \hline
\end{tabular}
\end{table}

Results are shown in Table \ref{tab:results}. For comparison, we also give the charged pion loop results obtained in the leading order calculation, HLS and ENJL approaches, and using the NLO operators in Eq.~(\ref{eq:leff1}) but imposing the cut-off $K^2< (500\ \mathrm{MeV})^2$ discussed earlier. As a cross check on our evaluation of the integrals, we have reproduced the LO and HLS results reported in Refs.~\cite{Kinoshita:1984it,Hayakawa:1995ps,Hayakawa:1996ki}. In the case of the ENJL model, one must include a momentum-dependence for the vector meson mass, an effect we are not able to implement using the integration procedure described above. However, taking a momentum-independent mass yields $-16\times 10^{-11}$, in good agreement with the full ENJL result reported in Ref. \cite{Bijnens:1995xf} . 

The last three lines in Table \ref{tab:results} include the results from Refs.~\cite{Bijnens:2012an,Abyaneh:2012ak} and the two models adopted in this work. The second and third columns give the results for two different values for $(\alpha_9^r+\alpha_{10}^r)$: (a) $(1.32\pm 1.4)\times 10^{-3}$, obtained using the results of pion radiative decay and (b) $(3.1\pm 0.9)\times 10^{-3}$, corresponding to the determination of the polarizability obtained from radiative pion photoproduction. Note that only the results in the last three lines of Table \ref{tab:results} depend on this choice. For case (a), the value of $M_A$ implied by Eq.~(\ref{eq:axial}) is about 20 \% larger than given by the KSFR relation; consequently, Model I no longer yields a finite value for $(\Delta m_\pi^2)_\mathrm{EM}$ for this choice.

Several features emerge from Table \ref{tab:results}:

\noindent (i) Inclusion of the polarizability tends to decrease the Standard Model prediction for $a_\mu^\mathrm{HLBL} $, regardless of which procedure one follows in treating its high momentum behavior, thereby increasing the discrepancy with the experimental result. 

\noindent (ii)Use of a model that interpolates to high momentum and that is consistent with the required asymptotic behavior of the virtual Compton amplitude leads to a substantially larger shift than does the imposition of a cut-off.
When compared to the Standard Model prediction obtained using the ENJL model, this shift can be has much as $\sim 30\times 10^{-11}$ 
[case (a)] or $\sim 60\times 10^{-11}$ [case (b)].
\noindent (iii) The uncertainties in the polarizability contribution associated with both the experimental value of $(\alpha_9^r+\alpha_{10}^r)$ and the choice of a model for interpolating to the asymptotic domain are significant, particularly compared with the anticipated experimental error for the future FNAL measurement of $\delta a_\mu = \pm 16\times 10^{-11}$.

Clearly, it will be desirable to reduce the uncertainties associated with polarizability contribution.  The planned JLab experiment will determine the polarizability LECs with an uncertainty of $\delta(\alpha_9^r+\alpha_{10}^r) = 0.16\times 10^{-3}$, thereby reducing the parametric error well below the level of the expected FNAL uncertainty  in $a_\mu$. Reducing the model-dependent uncertainty will require additional input. 
To be on the conservative side, one would like to have in hand an independent, experimental test of the momentum-dependence of the polarizability that could help discriminate between Models I and II and any other prescriptions for interpolating to the asymptotic domain. The possibilities for doing so will be the subject of forthcoming work.

\noindent \noindent{\it  Acknowledgements} 
We thank J. Donoghue, B. Holstein, and R. Miskimen for useful conversations; J. Bijnens for providing input on the NLO/cut-off evaluation; and D. Hertzog and D. Kawall for critical reading of this manuscript. 
The work is partially supported by U.S. Department of Energy contracts DE-FG02-92-ER40701 (KTE) and DE-FG02-08ER41531 (MJRM) and the Wisconsin Alumni Research Foundation (MJRM).


\end{document}